\documentclass[
reprint,
superscriptaddress,
 amsmath,amssymb,
 aps,
 prl
]{revtex4-2}

\usepackage{graphicx}
\usepackage{dcolumn}
\usepackage{bm}
\usepackage{hyperref}
\usepackage{xcolor}
\usepackage{amsmath}
\usepackage{cancel}
\usepackage{subfigure}
\usepackage[percent]{overpic}

\begin{document}

\title{Optical control of the moiré twist angle}
\author{Zhiren He}
\affiliation{Department of Physics, University of North Texas, Denton, TX 76203, USA}
\author{Prathap Kumar Jharapla}
\affiliation{Department of Physics, University of Seoul, Seoul 02504, Korea}
\author{Nicolas Leconte}
\affiliation{Department of Physics, University of Seoul, Seoul 02504, Korea}
\author{Jeil Jung}
\affiliation{Department of Physics, University of Seoul, Seoul 02504, Korea}
\author{Guru Khalsa}%
\affiliation{Department of Physics, University of North Texas, Denton, TX 76203, USA}
\date{\today}

\begin{abstract}
In this theoretical work, we propose an all-optical method for fast, precise manipulation of two-dimensional multilayers by transferring orbital angular momentum from phase-structured light (e.g. vortex beams) to a 2D material flake. We model the light-matter interaction, analyze the twist dynamics, and develop a phase diagram for optical twists by mapping the system onto an impulsively forced nonlinear pendulum. Our findings reveal rich dynamical responses spanning single- and multi-pulse twist angle control to (quasi)stable dynamical trajectories, and suggest a pathway for all-optical measurement of the twist potential energy. Aided by classical potential estimates for the interlayer energy and numerical simulation, we demonstrate the feasibility of this approach with hexagonal boron nitride bilayers and extend the results to dichalcogenides with first-principles calculations. These results can be generalized to other 2D multilayers, paving the way for scalable and customizable moiré electronics and photonics.
\end{abstract}

\maketitle

\textit{Introduction}\textemdash The sensitivity of the band structure and associated correlation effects to twist angle in two-dimensional 
(2D) materials offers a powerful handle for engineering novel quantum phases and functionalities \cite{Bistritzer_2011, Andrei_2021, Mak_2022, Du_2023}. However, this sensitivity also presents a significant challenge, as the exploration of electronic and optical properties relies heavily on the fabrication of high-quality moir\'e multilayers with precise control over twist angle. The earliest method of tear and stack resulted in random twist angles that were not tunable~\cite{Kim2016}. Various advanced assembly and twisting techniques have followed, including scanning microscopes \cite{Ribeiro_Palau_2018,Hu_2022,Yang_2020}, the cutting-rotation-stacking method \cite{Chen_2016}, mechanical bending \cite{Kapfer_2023}, and a recent approach based on an electrostatic microelectromechanical system (MEMS) \cite{Tang_2024}. Despite these advancements, these techniques still require sophisticated nano-fabrication/transfer processes, limiting scalability, reproducibility, and \textit{in-situ} tunability.

Spatially phase-structured light, such as a vortex beam, carries orbital angular momentum (OAM), providing an optical handle on the twist angle. Vortex beams have been used to rotate macroscopic objects like $\mu$m-sized beads or even three-dimensional optically trapped structures \cite{Paterson_2001, MacDonald_2002,Grier_2003,Shen_2019}. Vortex beams are traditionally generated in the near-infrared (IR) to visible range, where they primarily couple to electronic degrees of freedom. Recent advances extend their reach into the mid- and far-IR regimes~\cite{Araki_18,Tong_2020,Sharma_2022}. In parallel, recent development of powerful laser sources in mid- and far-IR frequency range have demonstrated significant potential for lattice and structural control \cite{Foerst2011,Mankowsky2017,Juraschek2017,Hortensius2020,Disa2021,Khalsa2024}. Although vortex beams have not been explored for this purpose, recent theoretical proposals suggest they can manipulate topological spin and polar textures \cite{Gao_2023,Gao_2024}, signaling that mid- and far-IR vortex beams could soon play a role in structural control of quantum materials and the manipulation of correlated phenomena. 

\begin{figure}[!ht]
\includegraphics[width=\linewidth]{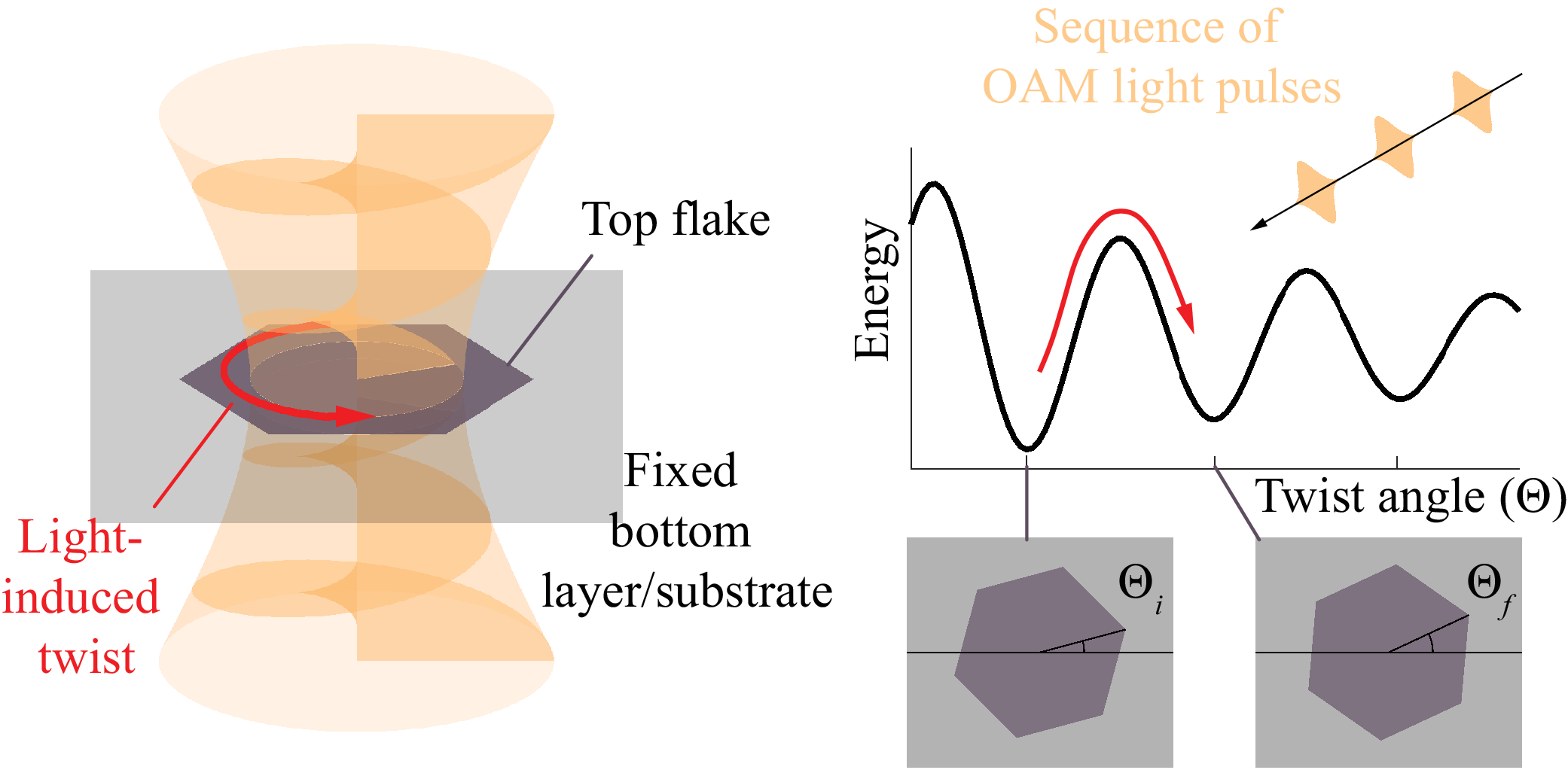}
\caption{\label{fig:schematic} Schematic of the optical twisted process. Left: A flake of 2D material situated on a surface/substrate gains angular momentum from vortex beam (orange). Right: The optically applied torque from a single, or a series of, optical pulses overcome a barrier in the twist potential energy, rotating the flake to a new metastable angle.
}
\end{figure}

In this work, we conceptualize an optical strategy for tuning the twist angle with OAM-carrying vortex beams in the mid- and far-IR, as shown in Fig. \ref{fig:schematic}. The OAM of light is efficiently converted to mechanical rotation of 2D flakes through resonant excitation of an IR-active phonon. Our strategy has the following key features: (1) pristine moir\'e multilayers can be directly manipulated without additional fabrication steps or direct contact, whether \emph{in situ} or \emph{in vacuo} and (2) a wide range of twist angles are accessible and tunable on a fast timescale. 

In what follows, we first illustrate the light-matter interaction in a general sense, showing that this strategy is applicable to many moir\'e homo/heterostructures. We then explore the twist dynamics of a 2D flake where we find rich dynamical responses and derive analytic results to guide experimental testing of this proposal. Then, using hexagonal boron nitride (hBN) as a case study, we combine pairwise classical potential results with numerical simulations to quantitatively validate the feasibility of our approach. Finally, the analysis is extended to moir\'e transition metal dichalcogenides (TMDs). 

\begin{figure}[!ht]
\centering
\subfigure{
    \begin{overpic}[width=\linewidth]{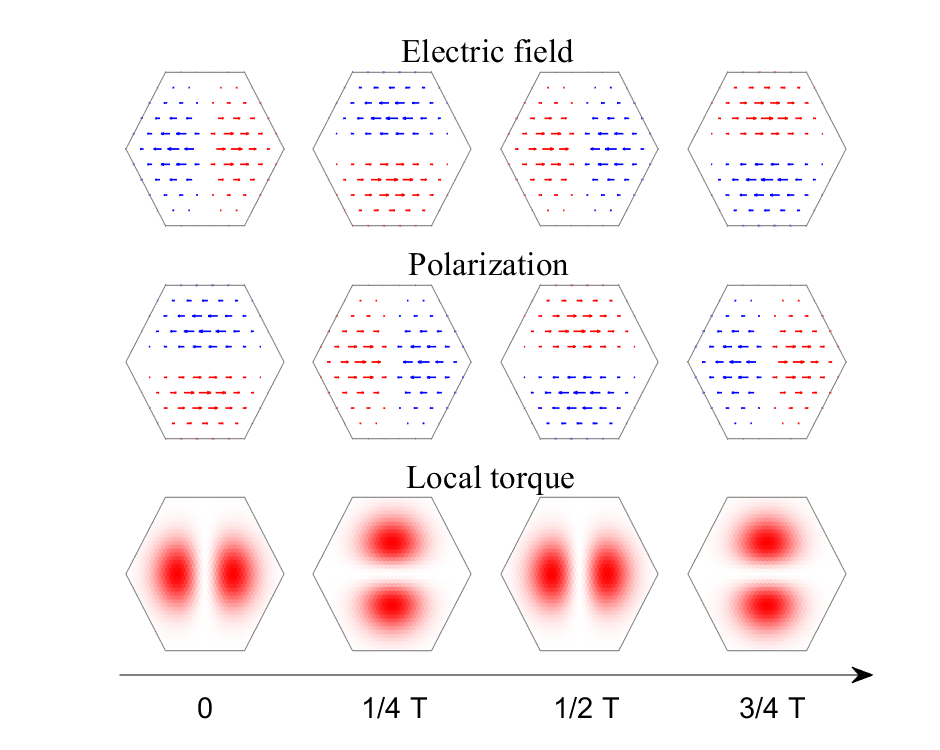}
        \put(0,75){(a)} 
    \end{overpic}
}
\subfigure{
    \begin{overpic}[width=\linewidth]{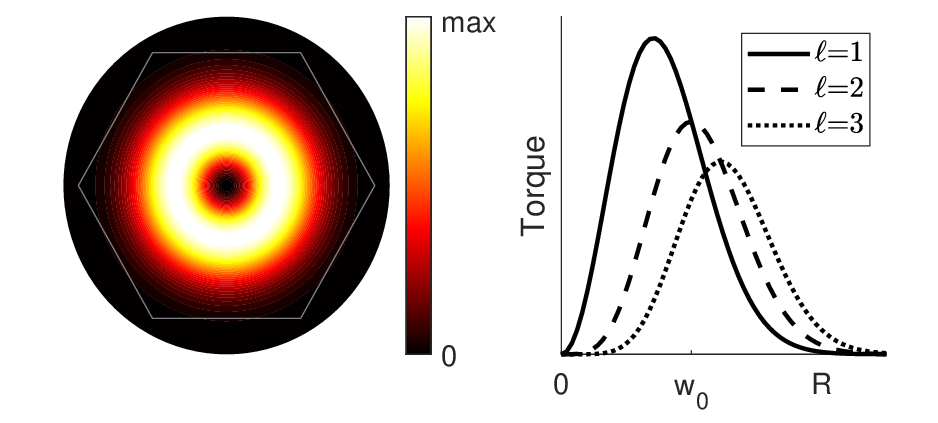}
        \put(0,45){(b)}
    \end{overpic}
}
\caption{\label{fig:fields} Torque generation via vortex beam-matter interaction. (a) Quarter-period snapshots of the distribution of electric field, electric polarization, and torque when a linearly polarized LG vortex beam ($\ell=1$, $p=0$, $w_0=0.5R$) resonantly excites an IR-active phonon in a hexagonal flake. Left and right electric field/polarization directions are represented by blue and red arrows, respectively, with lengths proportional to the magnitude. The local torque is positive everywhere, with torque magnitude proportional to the color saturation. (b) Left: Time-averaged torque distribution from (a). Right: Radial dependence of time-averaged torque for $\ell$ = 1, 2, 3, with $p=0$ and $w_0=0.5R$. $R$ is the distance from center to the vertex.}
\end{figure}%
\textit{Torque Generation with Phase-Structured Light} \textemdash  \ 
In circularly polarized light, \textit{spin angular momentum} (SAM) originates from the rotation of a uniform polarization plane. In contrast, orbital angular momentum arises when the phase of the wavefront is spatially structured. Here we briefly describe OAM transfer from a Laguerre-Gaussian (LG) beam to the lattice of stacked 2D materials, noting that the same principles apply to other types of phase-structured OAM light \textit{mutatis mutandis}.

To this end, suppose that a moir\'e multilayer is placed at the beam's focal point ($z=0$). The LG beam's spatial profile in cylindrical coordinates ($r,\phi,z$) takes the form
\begin{equation}
u(r,\phi,z=0)\propto \left(\frac{r \sqrt{2}}{w_0}\right)^{|\ell|}  L_p^{|\ell|}\left(\frac{2 r^2}{w_0^2}\right) e^{-r^2/w_0^2} e^{-i \ell \phi}
\end{equation}
$w_0$ is the beam waist radius and $L_p^{|\ell|} (x)$ is the associated Laguerre polynomial in $x$ of order $|\ell|$ and degree $p$. Here, OAM is indexed by $\ell$, derived from the phase structure $e^{-i \ell \phi}$, with positive/negative $\ell$ labeling clockwise/anti-clockwise rotation of beam profile. The index $p\ge0$ labels the number of radial nodes in the beam's spatial profile. A LG beam has a well-defined time-averaged total angular momentum density per unit power of \cite{Allen_1992}
\begin{equation}
    J = \frac{\ell}{\omega}|u|^2. \label{eqn:am}
\end{equation}

A vortex beam then transfers its angular momentum to matter via absorption processes. 
Although strong optical absorption can occur via interband electronic transitions,
we expect it to be an inefficient route in typical materials because this mechanism relies on the relatively weak electron-phonon coupling channels. Therefore, passing the angular momentum to the lattice \emph{directly} by exciting the IR-active phonon may prove to be a better strategy, especially near the IR resonance, where conversion to mechanical energy is maximized. 

A microscopic view of the angular momentum transfer process at resonance is shown in Fig. \ref{fig:fields}a for a 2D hexagonal insulating dielectric flake (no magnetization). The first row illustrates the time evolution of the spatial profile for $\ell=1$, $p=0$ LG beam. The nonuniform electric field of the LG beam induces a 2D polarization texture $\vec{P}(x,y)$ through the displacement of in-plane IR-active phonons $\vec{Q}_{\text{IR}}(x,y)$. When driven on resonance, the phase of the $\vec{Q}_{\text{IR}}$ and $\vec{P}$ will lag behind the electric field by a quarter-period, as seen in a resonantly driven damped harmonic oscillator. $\vec{P}$ creates bound charge ($\rho_b=-\nabla\cdot\vec{P}$) as well as bound current ($\vec{J}_b=\partial\vec{P}/\partial t$), both varying in time with the period of the electric field. The field-induced charge and current generate torque that acts directly on the lattice, as given by
\begin{equation} \label{eqn:local_torque}
    \vec{\tau}  = \int dA\ \vec{r}\times(\rho_b\vec{E}+\vec{J}_b\times\vec{B})
\end{equation}
\noindent
where the integral is over the area of the flake. The third row of Fig.\ref{fig:fields}a shows that the torque is not uniformly distributed, but positive everywhere on the surface and throughout the period.  The time-averaged torque profile, proportional to the intensity profile, has only radial dependence, which can be adjusted by beam characteristics through $\ell$, $p$, and $w_0$ (Fig. \ref{fig:fields}b). Further discussions of the transfer efficiency, circularly polarized LG beams, and edge effects are in Supplemental Information (S.I.) Section II, III, IV \cite{supplementary_info}. 

We now demonstrate that torque from a sequence of ultra-short pulses of vortex beams in the mid-/far-IR can overcome the twist energy barrier needed to alter the relative interlayer twist angle.

\textit{Modeling the Twist Dynamics}\textemdash  A single layer has access to a broad range of stable angles with the underlying multilayer or substrate, suggesting a corrugated potential energy landscape parameterized by twist angle $\Theta$ \cite{Zhu_2021}. Near any metastable reference twist angle, the potential energy landscape can be reasonably approximated as a sinusoidal function, 
\begin{equation} \label{eqn:twist_potential}
    U(\Theta) = \frac{1}{2}U_0\left[1- \cos(k\Theta)\right],
\end{equation}
which imposes a barrier of height $U_0$ to access neighboring twist angle minima. Here, $k$ parameterizes the local potential energy so that neighboring energy minima are separated by $\pm2\pi/k$. Displacing the angle $\Theta$ from equilibrium gives a restoring torque $\tau (\Theta) = -\partial_\Theta U(\Theta) = -\tau_0 \sin(\Theta)$, where $\tau_0 = k U_0/2$. The dynamics can thus be modeled in analogy to a rigid pendulum, with a moment of inertia $I$, and resonant frequency $\Omega_0=\sqrt{k^2U_0/2I}$ when $\Theta$ is small. Defining a rescaled coordinate $\Theta' = k\Theta$, we see that within this picture, access to a new twist angle is analogous to a rigid pendulum traversing its highest point, i.e. when $\Theta'$ crosses $\pm \pi$.

We thus model the twist dynamics with a damped-driven nonlinear pendulum equation,

\begin{equation} \label{eqn:eom}
    \frac{d^2\Theta'}{dt^2}+\gamma\frac{d\Theta'}{dt}+\Omega_0^2\sin \Theta'=\tau(t)/I
\end{equation}

\noindent
where $\gamma$ is the effective damping parameter and $\tau(t)$ is a driving torque which can be derived from Eq. \ref{eqn:am} or \ref{eqn:local_torque} (S.I. Section I, II, III \cite{supplementary_info}).

\begin{figure}[!ht]
\centering
\subfigure{
    \begin{overpic}[width=\linewidth]{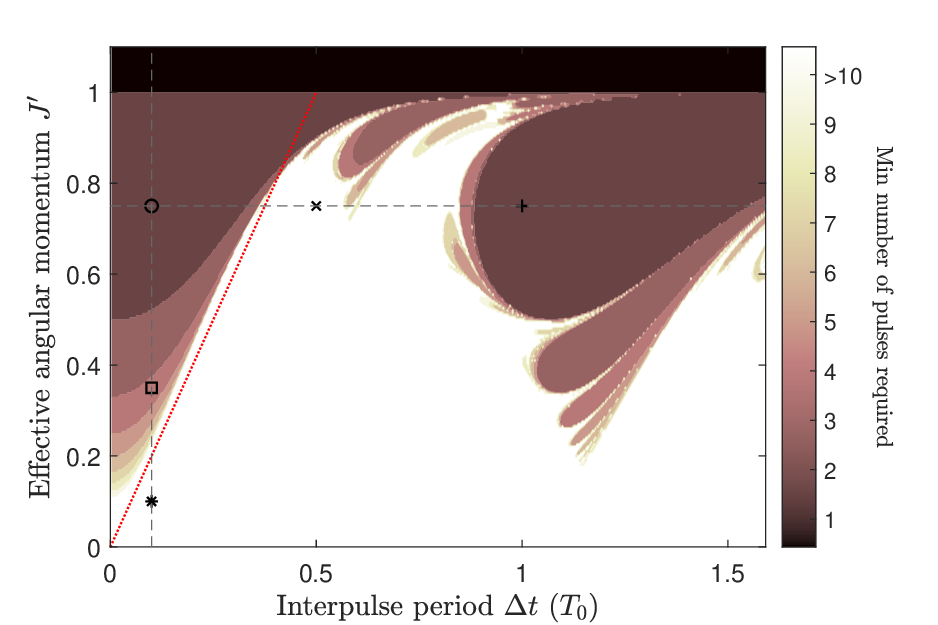}
        \put(0,60){(a)} 
    \end{overpic}
}
\subfigure{
    \begin{overpic}[width=\linewidth]{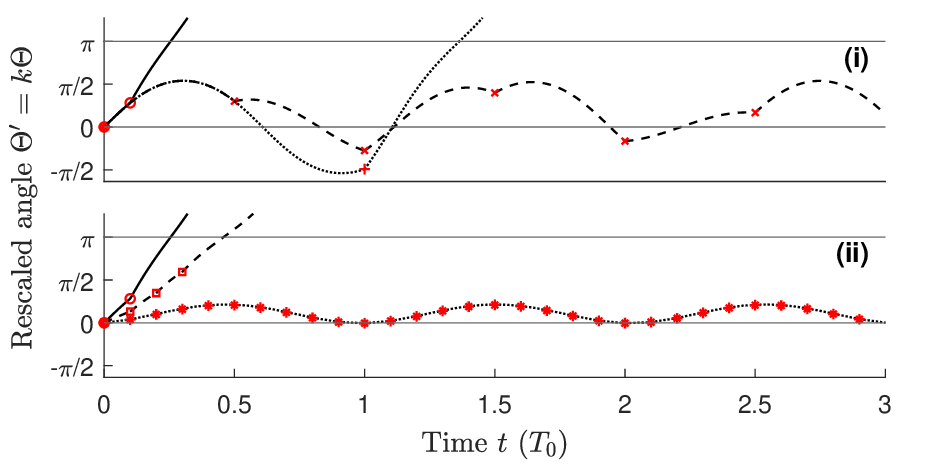}
        \put(0,50){(b)}
    \end{overpic}
}
\caption{\label{fig:dynamics}(a) Contour plot of the minimum number of pulses required to overcome a local twist potential energy barrier versus the effective angular momentum $J'$ and interpulse period $\Delta t$. When $J'>$ 1, a single pulse is enough to twist the flake. For $J' < 1$, complex dynamics emerge due to the intrinsic nonlinearity of the restoring torque. (b) Selected dynamics along horizontal line (i) and vertical line (ii) in (a).}
\end{figure}%
As the flake rotates about the center, its motion at the edge is limited by the speed of sound $v_s$. This suggests a lower bound for the twist angle period given by $T_0 = 2\pi/\Omega_0\sim v_s/R \approx 10$ ns, for a flake of radius $R\sim$100 $\mu$m and $v_s\sim 10^4\ $m/s. Conversely, mid-/far-IR laser pulses are typically prepared with 100 fs to 1 ps durations, generating optical torques that are reasonably approximated as \emph{impulsive},
\begin{equation} \label{eqn:torque}
\tau(t) = 2I\Omega_0 J'\sum_n{\delta(t-t_n)}.
\end{equation}
\noindent
Here $t_n$ is the arrival time of each LG pulse. The effective angular momentum $J'$ is the ratio of the angular momentum gained by the flake, $J_\text{optic}$, to that required to overcome the energy barrier, $J_0=2I\Omega_0/k$, expressed as
\begin{equation}
J' \equiv \frac{J_\text{optic}}{J_0}=\left[\frac{\ell+s}{\omega}\left(\Gamma_\text{abs}FA_\text{flake}\right)\right]/\left(\frac{2I\Omega_{0}}{k}\right), \label{eqn:fluence}
\end{equation} 
for each single pulse with fluence $F$, a flake with area $A_\text{flake}$, and absorption ratio $\Gamma_\text{abs}$. 

The impulsively driven dynamics described by Eq. \ref{eqn:eom} lead to rich dynamical behavior and the potential for chaos \cite{de_paula_chaos_2006}. Here, we focus our exploration on the conditions required to overcome the local potential barrier. 

We illustrate the twist dynamics for a sequence of evenly spaced LG pulses. We numerically integrate Eq.~\ref{eqn:eom}, tracking the minimum number of pulses required for the flake to overcome the twist energy barrier (i.e. when $\left|\Theta'\right|>\pi$). Fig. \ref{fig:dynamics}a shows this result with respect to $J'$ and the interpulse period $\Delta t\equiv(t_{n+1}- t_n)$ scaled by the natural period $T_0\equiv2\pi/\Omega_0$, in the limit of no damping. Including damping shifts the contour lines up in Fig. \ref{fig:dynamics}a, though the qualitative features remain unchanged (S.I. Section VI \cite{supplementary_info}).  This shift becomes more pronounced for longer $\Delta t$, since more kinetic energy will be dissipated.

We select several representative points in the phase diagram to illustrate the rich twist dynamics (Fig. \ref{fig:dynamics}b). When the repetition rate is high ($1/\Delta t\gg \Omega_0$), many impulsive torques are applied within a single period, producing a time-averaged torque of $2 I\Omega_0J'/\Delta t$. The need for this time-averaged drive to overcome the barrier energy defines a condition, $J'\ge 2 \Delta t/T_0$, which sets a bound on the laser characteristics required for switching (dotted red line of Fig. \ref{fig:dynamics}a), below which oscillatory motion about a shifted minimum is expected (`$\ast$'). The interval of potential energy curve with negative curvature causes a slight deviation from the dotted red line (S.I. Section VII).

As the interpulse period $\Delta t$ increases, we move through an out-of-phase drive region where switching requires fluence comparable to the single pulse threshold value. When the interpulse period is comparable to integer multiples of $T_0$, the periodic optical torque must work in concert with the angular velocity to overcome the restoring torque (`+'). However, without carefully timing the optical pulses, the system can remain in driven oscillatory motion for very long times (`$\times$'). More discussions on the rich dynamics are included in S.I. Section VII. 

With the general dynamics and the conditions for switching in hand, we now explore hBN as a representative candidate for optical control of moir\'e twist angle, illustrating the feasibility and limitations of this approach.

\textit{hBN case study}\textemdash Because hBN has a doubly-degenerate in-plane IR-active phonon at about 40 THz, it should be amenable to the optical twisting process. To assess the feasibility of our approach, we estimate $J'$ and $\Omega_0$ based on the intrinsic flake properties: potential energy variation with angle $U(\Theta)$ and the absorption ratio $\Gamma_\text{abs}$ ($\Gamma_\text{abs}$ discussed in S.I. Section II \cite{supplementary_info}). 

We calculate $U(\Theta)$ for rigid hexagonal hBN flakes with armchair edges of size $R/a_0$ ranging from 100 to 1000 on an extended and pinned layer of hBN following the recipe outlined in \cite{prathap2025_unpublished} (S.I. Section V). In our parameterization of $U(\Theta)$,  $\Theta=0$ corresponds to AA-stacking. Starting with AB-stacking inverts the extrema in energy, but does not affect the qualitative features (S.I. Section VI \cite{supplementary_info}).

\begin{figure}[!ht]
    \centering
\includegraphics[width=\linewidth]{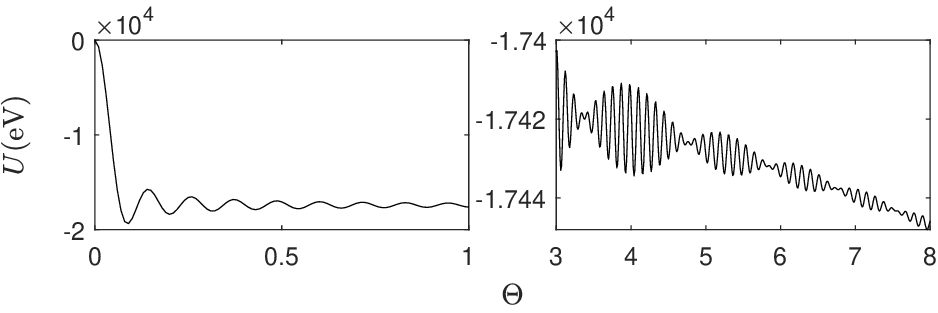}
    \caption{Total energy $U(\Theta)$ for flake size $R/a_0=500$. }
    \label{fig:beat}
\end{figure}

Fig. \ref{fig:beat} shows $U(\Theta)$ for flake size $R/a_0$ = 500. We identify a \emph{large-angle} region with beat patterns that begins near a critical angle of 3.3$^\circ$. The critical angle extrapolates to 0$^\circ$ as $R/a_0$ increases, making the large-angle region dominant for very large flakes (S.I. Section V \cite{supplementary_info}). In this large-angle region, when the potential energy is approximated locally with Eq. \ref{eqn:twist_potential}, the \emph{upper bound} on the barrier height within each beat is well described by $U_0(\Theta) = 2A\Theta^{-3}$, where $A\approx 748.3$ eV, independent of radius. Our fitting also shows that $k$ in Eq. \ref{eqn:twist_potential} is proportional to flake radius, but independent of the twist angle, i.e. $k=b R/a_0$, where $b\approx 0.1103$  (details of the scaling and fitting procedures in S.I. Section V \cite{supplementary_info}).

Using these results, the resonant frequency and effective angular momentum can be approximated as 
\begin{equation}
    \Omega_0=\sqrt{\frac{k^2U_0}{2I}}=\sqrt{\frac{ 2Ab^2}{\pi\rho a_0^2}}\frac{\Theta^{-3/2}}{R}
\end{equation}
\noindent
and
\begin{equation}
    J' = \frac{\ell+s}{\omega}\sqrt{\frac{\pi}{2\rho A}}\Gamma_\text{abs}F \Theta^{3/2},\label{eqn:fom}
\end{equation}
respectively, using a circular flake ($I=1/2\pi\rho R^4$) for simplicity.

Whereas directly simulating a flake size comparable with electronic and optical device fabrication needs (\textit{i.e.} $\sim 100\ \mu$m) is too cumbersome, our findings can be extrapolated from the simulation sizes to any desired scale. For a circular hBN flake of radius = 100 $\mu$m, absorption $\Gamma_\text{abs}=0.1$, IR-active phonon frequency $\omega=2\pi\times 40.3$ THz, $\ell=1$, $s=0$, and fluence $F=50$ mJ/cm$^2$, we find $J'=0.022 \ \Theta^{3/2}$. This gives the single-pulse switching boundary in Fig. \ref{fig:dynamics} (i.e. $J'=1$ ) at $\Theta \approx 12.7^\circ$. The flake's natural frequency is then $\Omega_0\approx 76.8\ \Theta^{-3/2}$ MHz, corresponding to an oscillation period of $T_0 \approx 81.8\ \Theta^{3/2}$ ns, comparable to our initial estimates. Above this angle, we anticipate single-pulse switching. Below this angle, a sequence of pulses will be necessary for moiré twist angle switching. Additionally, the dynamics portrayed in Fig. \ref{fig:dynamics} may be studied to map the twist potential. 
 
\textit{Discussion and Generalization} \textemdash The geometrical insight that the twist potential energy should depend on the relative areas of AA-stacked (high energy) and AB-stacked (low energy) regions enables an extension to other 2D materials without cumbersome computations ~\cite{prathap2025_unpublished}. We take the scaling relation $U_0(\Theta) = 2A\Theta^{-3}$ to be general, since it stems from the weak interlayer interaction, a general feature/necessary ingredient for 2D materials. We can then estimate the coefficient $A$ from the energy difference between the stacking configurations, i.e. $A\propto E(AA)-E(AB)$, a quantity readily found by DFT calculations (S.I. Section VII \cite{supplementary_info}). 

Under the same total angular momentum $\ell+s$, absorption coefficient $\Gamma_\text{abs}$, and twist angle $\Theta$, a comparison of $J'$ and $\Omega_{0}$ between hBN and MoX$_2$ becomes possible (TABLE \ref{tab:tmd}). Although MoX$_2$ flakes are heavier and exhibit larger twist barriers than hBN, they may be easier to twist. This is a consequence of the inverse proportionality of the OAM density to frequency $\omega$ (Eq. \ref{eqn:am}). For the same flake radius $R$, MoX$_2$ flakes also have lower natural frequencies $\Omega_{0}$. The combination of higher $J'$ and lower $\Omega_{0}$ suggests that the single-pulse switching boundary and tongue regions in Fig. \ref{fig:dynamics} are more accessible for MoX$_2$ flakes, provided similar pulse characteristics can be achieved at these pump frequencies.

\begin{table}[!htb]
    \centering
    \begin{ruledtabular}
    \begin{tabular}{ccccc}
    &  hBN & MoS$_2$ & MoSe$_2$ & MoTe$_2$\\ \hline
    $a_0$ (\AA) & 2.505 & 3.161 \cite{MoS2} & 3.288 \cite{MoSe2} & 3.519 \cite{MoTe2} \\
    $\rho$ (AMU/\AA$^2$) & 4.57 & 18.50 & 27.12 & 32.74\\
    $\omega/(2\pi)$ (THz) & 40.3 & 8.55, 11.53& 5.04, 8.60 & 3.56, 7.13 \\
    E(AA)-E(AB) (meV) & 27.2 & 72.5 & 82.6 & 113.0\\ \hline
    $J'$(MoX$_2$)/$J'$(hBN)& -- & 1.43, 1.06 & 1.88, 1.10 & 2.08, 1.04\\
    $\Omega_{0}$(MoX$_2$)/$\Omega_{0}$(hBN)& -- & 0.51 & 0.42 & 0.39\\
    \end{tabular}
    \end{ruledtabular}
    \caption{Estimation of the  effective angular momentum $J'$ and resonance frequency $\Omega_{0}$ for TMDs, compared with hBN. $a_0$ is the lattice constant, $\rho$ is the mass density, $\omega$ is the frequency of IR-active phonon in the plane. For MoX$_2$, there are 2 in-plane IR-active phonons. $E(AA)-E(AB)$ is the energy difference between the AA and AB stacking order.}
    \label{tab:tmd}
\end{table}%
A successful optical twist may require less power than estimated in this work due to our rigid-flake approximation and omission of thermal effect in the modeling of $U(\Theta)$, further discussed in S.I. Section IX. On the other hand, in the non-switching scenario, since OAM light can excite the rotational degrees of freedom, direct optical study and measurement of the twist potential and fast (sub-nanosecond) twist dynamics become possible. This opens up new avenues for exploration of the fundamental physics of angular momentum transfer and relaxation dynamics in 2D materials, though resolving the micro-/mesoscopic structural changes in this broad temporal range will require development on both theoretical and experimental fronts. 

\textit{Conclusion}\textemdash We present an optical strategy for fast, precise, and potentially scalable control of the moiré twist angle in 2D multilayers through the use of OAM-carrying vortex beams in the mid- and far-IR. The key physical realization is that OAM pulses can impart substantial angular momentum on large-scale 2D materials flakes through the excitation of IR-active phonons. This induces complex dynamics and switching, provided the interlayer potential barrier can be overcome. Using the example of hBN, we find that current laser sources are deployable to initiate the study of this strategy for moiré twist-angle control. Finally, this approach is readily generalized to a broad class of 2D materials.

\textit{Acknowledgments}\textemdash 
This work was supported by the Korean NRF through Grant NRF-2020R1A5A1016518 (P.K.J. and J.J.) and Grant RS-2023-00249414 (N.L.). We acknowledge computational support from KISTI Grant No. KSC-2022-CRE-0514 and by the resources of Urban Big data and AI Institute (UBAI) at UOS, and Texas Advanced Computing Center (TACC) at the University of Texas at Austin.

\bibliography{ref}
\end{document}